\documentclass{article}

\usepackage{PRIMEarxiv}

\usepackage[utf8]{inputenc} 
\usepackage[T1]{fontenc}    
\usepackage{hyperref}       
\usepackage{url}            
\usepackage{booktabs}       
\usepackage{amsfonts}       
\usepackage{nicefrac}       
\usepackage{microtype}      
\usepackage{lipsum}
\usepackage{fancyhdr}       
\usepackage{graphicx}       
\usepackage{hyperref}
\usepackage{xcolor}

\usepackage{geometry}
\usepackage{cleveref}
\usepackage{nameref}

\usepackage[backend=biber,style=apa, url=false]{biblatex}
\title{Generative Adversarial Collaborations:\\ A practical guide for conference organizers and participating scientists}

\author{
  Gunnar Blohm\textsuperscript{*} \\
  Centre for Neuroscience Studies \\
  Queen’s University \\
  Kingston, Ontario, Canada \\
  \And
  Benjamin Peters\textsuperscript{*} \\
  School of Psychology \& Neuroscience \\
  University of Glasgow \\
  Glasgow, UK \\
  \And
  Ralf Haefner \\
  Brain and Cognitive Sciences \\ 
  University of Rochester \\ 
  Rochester, USA \\
  \And
  Leyla Isik \\
  Department of Cognitive Science \\
  Johns Hopkins University \\ 
  Baltimore, Maryland, USA \\
  \And
  Nikolaus Kriegeskorte\\
  Zuckerman Mind Brain Behavior Institute \\ 
  Columbia University \\ 
  New York, New York, USA \\  
  \And
  Jennifer S. Lieberman\\
  Zuckerman Mind Brain Behavior Institute \\ 
  Columbia University \\ 
  New York, New York, USA \\    
  \And
  Carlos R. Ponce \\
  Department of Neurobiology \\ 
  Harvard Medical School \\ 
  Boston, Massachusetts, USA \\
  \And
  Gemma Roig \\
  Department of Computer Science \\ 
  Goethe Frankfurt University \\ 
  Frankfurt a.M., Germany \\
  \And
  Megan A. K. Peters\textsuperscript{†}  \\
  Department of Cognitive Sciences \\
  University of California Irvine \\ 
  Irvine, California, USA \\
}

\begin{document}

\def\thefootnote{*}\footnotetext{These authors contributed equally to this work.}
\def\thefootnote{†}\footnotetext{Correspondence should be addressed to Megan A. K. Peters, \href{mailto:megan.peters@uci.edu}{megan.peters@uci.edu}}

\maketitle

\begin{abstract}
Generative adversarial collaborations (GACs) are a form of formal teamwork between groups of scientists with diverging views. The goal of GACs is to identify and ultimately resolve the most important challenges, controversies, and exciting theoretical and empirical debates in a given research field. A GAC team would develop specific, agreed-upon avenues to resolve debates in order to move a field of research forward in a collaborative way. Such adversarial collaborations have many benefits and opportunities but also come with challenges. Here, we use our experience from (1) creating and running the GAC program for the Cognitive Computational Neuroscience (CCN) conference and (2) implementing and leading GACs on particular scientific problems to provide a practical guide for future GAC program organizers and leaders of individual GACs.
\end{abstract}


\section*{Introduction}
We commonly observe research groups within a given field, each defending their own favorite framework, theory, or model. Most science proceeds with competing viewpoints investigated and published in parallel, following separate lines of inquiry, and by separate groups (\href{https://gac.ccneuro.org/call-for-proposals}{https://gac.ccneuro.org/call-for-proposals}). This leads to a situation where research efforts will be directed at confirming said theories while invalidating competing viewpoints. Such an approach to research results in a polarized field, where, for example, opposing groups will argue that competing groups didn’t carry out the right experiment or that a key control condition is missing. We often hear researchers claim that competition drives innovation and advances. While there is certainly truth to that, polarized competition may often lead to siloed research fields and rarely results in the resolution of debate that would significantly advance science in a meaningful way. Even if it does lead to a resolution of controversy, this typically takes a very long time. Naturally, the question arises: is there a more efficient way to advance science?

This is where Generative Adversarial Collaborations (GACs) come in. GACs are collaborative endeavors specifically designed around bringing opposing camps with diverging theories and viewpoints together in an honest attempt to identify a path toward resolving the controversies that arose from the opposing viewpoints. Inspired by previous observations about the benefits of such ``adversarial collaborations'' \parencite{kahneman2003experiences, latham1988resolving}, GACs aim to foster direct debates that lead to collaboration. They invite researchers to discuss their opposing viewpoints in a constructive and amicable way. The goal of GACs is thus to work out the exact nature of disagreements and devise an agreed-upon plan to resolve them. The more general goal for each GAC is to accelerate advances in science by focusing on finding solutions in a collaborative way instead of perpetuating the cycle of trying to validate one’s own pet theory. Ideally, the GAC process should be open to contributions from the research field at large; the more people contribute to finding (and executing) the best way forward, the better the chances of success. 

Here, we describe what we have learned from the GAC endeavors spearheaded at the Cognitive Computational Neuroscience (CCN) conference over the past four years, summarizing experiences and feedback from 15 GAC teams so far. As GAC program organizers, we have followed the progress of the GAC teams and, in 2021, held a workshop to discuss what worked, what was challenging and how to improve the process. Subsequently, we have gathered feedback from both GAC teams and our broader community. As a result, we discuss here the benefits, challenges and suggested best practices for GAC teams in the hope that more researchers will engage in these collaborative endeavors. 

\section*{Ingredients of GACs}

The CCN program committee designed GACs to be a multi-step process. First, open calls for submissions of core teams undergo review by the GAC program committee to ensure appropriateness and clarity of the controversies at hand. Next, all proposals are published for open review to allow the research community to engage, provide feedback, and contribute to the GAC team’s ideas and plans -- or potentially even join the team. The GAC teams are then encouraged to take comments and contributions into account in the organization of a kick-off workshop hosted by CCN. The kickoff workshops are also an opportunity to expand the GAC team.

CCN GAC workshops were highly interactive, starting with an overview of the question and a debate between two or more proponents of competing theories and then moving to commentaries and reactions from the rest of the GAC team. GAC kickoff discussions were approximately 1-2 hours long, followed by structured commentary, fireside chat, and/or other formats presented by other GAC and community members for the remainder of the 3-4-hour events. Importantly, we expected these CCN GACs not to be a series of isolated talks, but instead an integrated presentation of competing viewpoints and open questions focused on moving the scientific process forward. Key components of the debate and discussion included a focus on identifying conflicting experimental predictions and results and brainstorming specific experimental approaches and potential empirical (or theoretical) outcomes that might arbitrate the competing theories. Much emphasis was put on sufficient time for discussion that included the audience. 

After the GAC kick-off workshop, GAC teams were encouraged to continue their discussion (we created a Slack channel for them to facilitate this) and to formalize their devised avenues forward in collaborative position papers, for which we created a special issue in collaboration with the Neurons, Behavior, Data, and Theory journal (NBDT, a platinum open-access journal) (\href{https://nbdt.scholasticahq.com/section/2529-generative-adversarial-collaborations-ccn-gacs}{https://nbdt.scholasticahq.com/section/2529-generative-adversarial-collaborations-ccn-gacs}). The idea of the position paper is to allow the research community to follow the logic of the controversies at hand and also learn about suggested ways forward. We also invited contributed commentaries for these position papers from the community. Ultimately, we hope that this process will allow other researchers to follow the proposed research advancements and potentially contribute to them through active research, effectively broadening the reach and impact of the GACs on their respective research communities. As such, GACs are meant to be medium to long-term collaborative projects. 

GAC proposals, recordings of the kick-off events, and position papers (where applicable) can be found here: \href{https://gac.ccneuro.org/gacs-by-year}{https://gac.ccneuro.org/gacs-by-year}. 

\section*{Benefits of GACs}

\subsection*{Conceptual benefits to the project from the ‘adversarial’ nature of the collaboration}

A GAC is, by definition, a collection of scientists who fundamentally disagree on at least one aspect of the topic at hand. What benefits can we expect to the actual project or scientific inquiry, due to this conceptual opposition? 

\subsubsection*{Science progresses through comparison of models and theories}

GACs formalize the practice of ``doing science'' as a form of comparison of models and theories. In many empirical sciences like cognitive and computational neuroscience, theories and models undergo continuous revisions as they are faced with new data. A more productive question than ``is this hypothesis, model, or theory correct?'' is therefore: ``which of these hypotheses, models, or theories better explains the data?'' Yet despite our best intentions, we as researchers are prone to insidious confirmation biases: it is easy (and pleasurable) to integrate new empirical findings directly into our favorite theory rather than considering them as evidence for alternative perspectives. Such confirmation biases have real-world consequences beyond the researcher’s mind, because they also inform the design of new experiments or theoretical approaches in support of the ‘favorite’ theory -- thus perpetuating isolated silos of research practices, approaches, and social groups of researchers.

GACs, not unlike formal model comparisons, change this dynamic by providing a forum where competing interpretations of the same empirical result must be actively discussed. This approach has the additional advantage of requiring proponents of each perspective to explicitly state any auxiliary assumptions and their commitments to the domain of a given model or theory, i.e., the range of observations that it is claimed to explain. During the GAC process, many teams discover that different interpretations of the same empirical evidence might be taken as support for diametrically opposed theoretical positions. This observation is illuminating for two reasons. First, it brings into focus just how common it is for us to avoid attempting to fit new results into a competing theory, both because we would like to be right and because we may not know the intricacies of opposing theories as comprehensively as we know our own. Simply recognizing the harm to the scientific process caused by confirmation biases and the ensuing intellectual silos is an important step. Second, recognizing how even the most elegant empirical results can be interpreted in multiple ways inevitably drives the development of new experimental or theoretical approaches that are specifically designed to arbitrate between competing viewpoints. The design of such crucial experiments \parencite{lakatos1974role, bacon1878novum} can often exceed the capacity or expertise of a single research group, but GACs pool the expertise across many groups to overcome this limitation.

\subsubsection*{Creating a common vocabulary}

One of the most common themes discussed in the GACs is the observation that at least part of the controversy may stem from different understandings or definitions of the same terms or models. For example, in initial conversations, a GAC team might discover that a core word like ``generalization'' or ``representation'' takes on fundamentally different meanings on the different ``sides'' of the controversy, and that the disagreement can be largely attributed to what amounts to a translation problem. But those seem to be the rarer cases. Another possibility is that, after longer discussions focused on aligning such definitions, progress can only be made if the group decomposes these key terms altogether, breaking down the concepts into more precise, operationalized definitions. Sometimes, that factorization process results in the resolution of parts of the disagreement, but even if this is not the result, discovering shared vocabularies and precise definitions can drive much more fruitful conversations going forward. 

The extent of this linguistic vagueness -- let alone the possibility of resolution -- is much more difficult to achieve in asynchronous communication that occurs primarily through sequential peer-reviewed publication or conference proceedings than in a focused workshop dedicated precisely to revealing such points of contention. The GAC process, including the position papers that arise from the workshop and ongoing collaborations, creates a shared vocabulary to anchor the team and help them develop concrete or practical steps to help progress their field.  

\subsubsection*{Eliminating misunderstandings}

Both of the previously discussed benefits -- actively seeking divergent views and creating a common vocabulary -- contribute to the broader benefit of eliminating misunderstandings between opposing theories. Proponents of one theory might strongly oppose another simply because they have misunderstood a core element, even if they are experts in the field and even after terminology misalignments have been resolved. Taking the time to discuss and resolve these misunderstandings in great detail can feel like a luxury; the GAC structure creates a space where resolving such misunderstandings is a necessity, so that the core of the controversy can be distilled and made practically accessible through pragmatic, tractable goals.

\subsubsection*{Discovering more agreement than expected}

Finally, as mentioned above, even beyond aligning vocabularies, categorizing empirical results, and eliminating misunderstandings, the GAC process can provide additional benefits. That is, once the air has been cleared, the ``opposing'' viewpoints might not be so opposing after all. Some of the GAC teams have discovered that their controversy largely dissipated (e.g. it was due to misalignment of vocabulary rather than conceptual disagreement); others have discovered a new opposing viewpoint that both sides can agree is a common ``foe''. In these and other possible outcomes, the focus of the GACs on finding a common path forward -- rather than on winning a debate -- creates a shared foundation that enables real progress instead of remaining stuck in an endless loop of miscommunications and biased ``confirmations''. It is more fun to work as a team than to work against one another.

\subsection*{Tangible benefits to both project and community from the GAC mechanism}

The GAC mechanism provides a formal structure for achieving the benefits listed above, which can greatly improve an individual project once a team is formed. In addition, GACs have interesting benefits for the scientific community and fields more broadly. 

\subsubsection*{Community engagement at multiple stages}

As described previously, the typical science life cycle can be narrow and local at the early stages of a project: small groups of collaborators, or even just an individual faculty member and their trainee, will develop a project with limited feedback from the broader community. This occurs for several reasons, including fear of being scooped or fear of appearing ignorant of core knowledge and developments in the field. This isolation is actively damaging to the quality of the project: Who among us has not given a talk on a nearly-finished project, only to receive a question from the audience that makes us think, ``I really wish I’d known that 6 months ago''? 

The GAC mechanism opens the discussion, and the focus on identifying practical paths forward, from the very beginning. From the moment a GAC is selected for a kickoff workshop, their proposal is posted on an open forum and community feedback is solicited. The goal of this process is twofold. First, the process can crowdsource important contributions to the project: empirical results, theoretical positions, omitted knowledge, and other factors that should play a strong role in the ensuing discussion. Second, opening up the proposals to the community provides an explicit opportunity for other researchers to get involved with a GAC if they wish to. Of course, there can be many reasons why a researcher would choose not to formally engage with a planned GAC; this is why there is a later stage of potential interaction, i.e., once a position paper is written, the community can engage in the discussion of this publication through writing commentary papers. The goal is to always keep avenues for community feedback, and to keep a track record of the GAC project’s evolution throughout time, so that researchers new to the field can catch up to the debate. 

\subsubsection*{Continuing systematic discussions}

One impediment to collaboration with a larger group of researchers is that it is hard to coordinate and commit to regular systematic discussions. The GAC mechanism is designed to facilitate this process by providing a platform (CCN kickoff workshop, Slack channel), setting goals (position paper) and following up with GAC teams (CCN GAC progress and process discussion, Slack follow-up). In doing so, GACs set the stage for deeper discussions. This is particularly empowering in the context of entrenched, potentially multi-generational debates, such as the question of whether the brain carries out probabilistic computations. Over the past four GAC cycles, these ongoing (online) discussions have turned out to be much more valuable than a single long in-person workshop or event. This is because cumulatively, over time, debates deepen. Participants have time to cogitate for days or weeks between discussions. This cannot be achieved at a single event. There is real value in this iterative process of meeting and discussing, taking a break to think about what has been discussed and then coming together again, over and over. Progress in thinking is indeed – just like learning – an iterative process. GACs fully take advantage of this iterative deepening of the discussion. 

\subsubsection*{Multi-generational discussions and mentorship}

GACs provide a real learning opportunity for researchers at all career stages. More senior researchers can learn new techniques and approaches from more junior (typically more hands-on) researchers. In return, junior members can benefit from the perspectives and mentorship of established contributors. Since GACs by design are open, anyone can join; indeed, the majority of GACs thus far have included a mix of senior faculty, early-career researchers, postdocs, and students. As such, GACs indirectly facilitate training and mentoring in the field in a new way that is not restricted to a specific lab or location. Instead, they provide networking opportunities worldwide, which is particularly interesting for researchers in less privileged situations who might lack access to good mentorship and/or networking opportunities. 

The open nature of GACs also means that researchers can, in principle, join in at any time. This means that GAC teams are dynamic, which can offer a number of benefits. One can imagine that the original team slowly hands over leadership to a newer generation. As thinking evolves, so might the composition of the team, opening to new expertise and/or viewpoints. Senior researchers over the years might retire or change their research focus. Young researchers will join the field and get excited to contribute. As such, GACs are meant to be a living, evolving environment for making scientific progress, networking, and mentorship alike -- and one that is not restricted to a lab, institution, country, continent, or time zone. 

\subsubsection*{Keeping the fun in science}

Most researchers would agree that discussing science is a lot of fun. GACs are meant to promote this discussion on several levels. But most importantly, GACs are meant to focus the discussion on progress and agreement despite divergent viewpoints of team members, instead of focusing on the disagreements and adversarial attitudes that so often poison our fields. Clarification of disagreement in an amicable, collaborative way really can lead to fun talks, discussions, and debates at several scales. First, of course, these discussions occur within the core GAC team. Second, the GAC kickoff workshops over the past two years were perceived as some of the most fun events ever for program organizers, GAC leaders and teams, and audiences. Presenting consecutive talks with opposing views generated tremendously fun discussions and debates. Participants of these events expressed the wish that more such debates should happen in science. Third, community involvement and feedback along all stages of the GAC process, from proposal to kickoff to opinion papers and beyond, provide fun opportunities for discussion and input. Complementary outside views are always refreshing and often allow recentering or refocusing of the debate. But most importantly, the more contributions, the more fun a GAC project can be. 

\section*{Challenges of GACs}

Clearly, science and scientific collaborations also have their challenges -- and GACs are no exception. Here we summarize 9 frequent challenges that CCN GAC teams have encountered and provide potential avenues to overcome those challenges and turn them into positive aspects of the GAC.

\subsection*{Scientific/collaboration challenges}

\textbf{Challenge 1: It is difficult to narrow the scope of the debate without getting lost in rabbit holes.} This is, of course, true for all scientific endeavors. However, GACs might be particularly vulnerable to getting lost in the details and/or deviating from the original goals, since their purpose is largely centered on identifying and exploring wedge issues and points of contention. However, there is also an opportunity to collectively explore aspects of the topic at hand that individuals have not thought of or that have traditionally been neglected in the field. It is an opportunity to reach new depths of thinking about topics; explore novel territory that was originally out of scope, or build new bridges between topics, ideas, and approaches. One path is to explicitly limit time for certain detail-focused discussions. Alternatively, sub-groups could engage in those discussions and summarize their conclusions to the rest of the GAC. Finally, we believe that deep debates are ultimately crucial for science. It is less important to find immediate answers to deep questions than to lay out the discussion to the community so that others can engage and contribute. This should be part of the goal of GAC position papers and accompanying commentaries. 

\textbf{Challenge 2: It’s not always easy to change views or even find common ground or compromise.} GACs are all about fruitful discussion. It can appear very difficult for opposing camps to find points of agreement, and the inability to bring others around to your viewpoint can be highly frustrating. The harder this seems, the more there is a need for discussion. The less common ground, the more that is a sign that a community is deeply divided. Consequently, this is really where GACs are sorely needed -- especially when viewpoints seem irreconcilably disparate. They offer the opportunity to rebuild trust, to re-establish communication, and importantly to rebuild common ground: Surely, there are many things that everyone participating in the GAC could agree upon, despite diametrically opposing views on one core issue. If not, then there is even more so a need to talk things through. Overcoming one’s own biases and beliefs and considering new viewpoints when they are deeply ingrained is difficult. But this is ultimately how science makes progress: not by insisting on being right, but by being open to alternatives, being curious to understand different viewpoints (and their potential benefits and flaws), and accepting that one might be wrong. Clearly communicating the agreements and disagreements, including their arguments, is part of what GACs can do to better inform their field and initiate deeper thinking about the contentious topics. 

\textbf{Challenge 3: The team needs goodwill to sustain the dialogue through many arguments.} This might seem trivial, but it is a crucial point. GAC members need to be open-minded and curious about understanding how and why opposing views came about. They have to be vulnerable in their own perspective and allow for doubt and questioning. They have to be willing to engage in a rational, objective, scientific discussion without attribution of personal blame or accusations of dishonesty. We are all passionate about science and it may be frustrating when our points of view are not respected and/or we feel we are not allowed to explain why we think differently. Dialogue -- as the name says -- goes both ways. Both parties have to be open, and maintaining this openness can be highly challenging, especially when one’s core scientific view is being challenged. Yet the benefit of maintaining the belief that all GAC members are acting in good faith is tremendous: doing so creates opportunities to move the field forward, to end up both being right and wrong, and to emerge from challenging discussions as winners. Because a successful GAC means the whole field wins. For that to happen, a continued open, honest dialogue with good intentions is needed. It helps to explicitly keep the end goal in mind.

\textbf{Challenge 4: We do not use the same words in the same way, even if we intend to. Language can be imprecise.} How many times have we been in a heated discussion only to find out that our arguments resulted from us having different definitions of words? Rather than being purely a challenge, this is a central opportunity for GACs to clearly identify such words of contention and to actively work towards a proper definition that everyone can agree upon, ensuring that the definition of words does not become the main argument. Again, teams are encouraged to communicate these definitions and agreements on meanings to the rest of the community -- ideally in a GAC position paper. 

\textbf{Challenge 5: It can be hard to find a middle ground without ``watering down'' the debate, especially in writing.} GACs are not all about agreeing on all aspects, but about finding positions that allow the field to move forward in a constructive way. It is thus important to provide details of arguments, especially if there does not seem to be an easy middle ground. Science does not benefit from vague, over-generalized statements. Rather, this situation should be seen as an opportunity to dig deeper and lay out the differences as substantively as possible. Writing down arguments is also important here as writing forces logical flow. Often, having to express ideas or arguments in writing highlights the inconsistencies or gaps in our logic. Use disagreement as an opportunity to precisely spell out differences and causes of differences between points of view. 

\textbf{Challenge 6: Facilitating and supporting ongoing community engagement is challenging.} Science best makes progress when engaging the whole community: Together, not as individuals, we drive science forward. As such, research itself is a community endeavor. GACs that attempt to resolve important controversies and challenges in the field should thus be open to ongoing community engagement. However, practically this can be difficult to achieve. We are all busy, and effectively engaging the community requires time, organization, and effort. The CCN GAC process attempts to help facilitate community engagement, at least early on in the process, so that the logistical and practical challenges of community involvement can be offloaded from the GAC members themselves. To this end, CCN invites all CCN members to comment on GAC proposals in an open review format; GAC kick-off workshops at the CCN annual meeting emphasize community contributions and discussions. Furthermore, GAC leaders are explicitly encouraged to invite community members to contribute to the discussion, formally or informally. However, once a GAC project has kicked off, it is essentially up to the GAC leaders and teams to continue engaging the community as each GAC has unique needs. While this can be done through the GAC position papers (and call for commentaries, also facilitated by CCN), there are other potential avenues, such as: GAC progress presentations at conferences, standing invitations to asynchronous discussion groups (by email, Slack, Discord, or other), publication of progress reports, or organization of focused workshops or conference sessions. Ultimately, it is up to each group/field to decide how to best move forward as a community for the benefit of their debate and the field as a whole. 

\subsection*{Practical Challenges}

\textbf{Challenge 7: GAC team members come to the project with different levels of capacity and willingness to engage.} Like in every collaborative project, individual contributions vary widely due to various constraints, interests, and priorities. The ideal pace for making progress is a weekly (or bi-weekly) standing meeting, but carving out this time can be difficult to justify for some. As a result, some GAC members are more present and available for discussions and driving the project forward than others. One opportunity resulting from the need to update members who cannot attend all meetings is to get into the habit of keeping detailed meeting notes. The added benefit of such notes is that they spell out arguments and discussions and can as such provide the basis for opinion papers, grant applications, etc. One mitigation approach is to appoint a project ``driver'' to maintain momentum -- often an early career researchers (ECR) who is the first author of the ensuing GAC position paper. Another possible mitigation strategy would be the addition of financial motivation, i.e., ideally (even just modest) funding for collaborative GAC projects; seed grants or pilot awards would help keep people accountable and would contribute to sustaining momentum. Below, we expand on these ideas in the \nameref{sec:incentive} and \nameref{sec:well-forward} sections.

\textbf{Challenge 8: Making progress only remotely over Zoom calls can be slow.} Meeting regularly online has many benefits, from accessibility to low-cost/high-frequency of meetings that can be spread out over time. However, setting up such meetings across broad time zones world-wide can be challenging, and ``Zoom fatigue'' makes videoconferencing a non-preferred working mode for many. One potential solution here is to use in-person conferences as opportunities to organize sessions around the GAC, or to have a satellite GAC workshop for longer, in-person discussions. However, as emphasized above, one-off in-person meetings should supplement, not replace, more regular, frequent online meetings and discussions. 

\textbf{Challenge 9: ``Where do we go from here?''} This question was often asked by GACs. At CCN, we purposefully left future steps open because we envisioned every GAC team wanting to work out on their own what avenues forward would be best for them and which goals they would want to set for themselves. Questions around the continuation of the GAC after the position paper were: Should we apply for funding? Should we write more papers? How can we integrate GACs into the traditional science credit system? How can we incentivize the continuation of the GAC project? While, as GAC program organizers, we don’t have a definite answer to these questions, we outline below some potential avenues forward to ensure the success of new GACs and the continuing success of ongoing GACs. 

\subsection*{Incentive structures}\label{sec:incentive}

GAC teams might find themselves ``stuck'' after the initial kickoff event, due to misaligned incentives and priorities. While it should be left to each GAC team to decide how to proceed after the kickoff event, we strongly encourage a continuation and at least a position paper as a written conclusion. Synthesizing insights from discussions had on Zoom and during the kickoff workshop in a paper is key for lasting community impact. And it is often the writing process that yields the kind of clarity and novel scientific insights one cannot obtain through discussions alone.

The main incentive for participating in a GAC is to make theoretical progress on key scientific problems. GACs can also be an opportunity to apply for funding and to be an author on an important publication for the field. GAC members will also have ``softer'' motivations for participating, e.g., enjoying intellectually stimulating discussions, building new collaborations, extending scientific networks, or increasing their visibility as experts in the community. Many of these motivations may have already been fulfilled with the conclusion of the GAC kickoff event. This is particularly true for senior members, who don't necessarily ``need'' a publication and for which continued Zoom meetings and collaborative writing adds another item to an already overfull agenda.

Acknowledging and refining the incentive structure will help counteract getting stuck in the GAC project. We discuss measures on reward structure, deadlines, and involving grant agencies and journal editors below.

\section*{How to do GACs well going forward}\label{sec:well-forward}

\subsection*{5 simple rules for successful GAC teams}

\textbf{Rule 1: Optimize group time.} We’re all busy. GACs are most successful when time is spent effectively. Regular team meetings are important to keep engagement and motivation. But no one wants to meet for the sake of meeting. Make sure you have a plan (agenda) for each meeting outlining specific goals, discussion points, questions, etc. Make sure agendas are reasonable and goals are achievable. Share agendas in advance to allow team members to contribute to them and prepare for the meeting. Collectively take detailed meeting notes in a single shared document for those team members unable to attend. Meeting notes will also be a written trace of the evolution of your thinking and debate, which comes in handy when trying to write things up. 

\textbf{Rule 2: Clarify your project goals.} Choose specific goals for your GAC. Start with small, attainable goals, ideally with a concrete output (e.g. publication) at the end to keep sharing with the rest of the community and maintain a trace of real progress. Narrow scopes ensure the feasibility of the project and guarantee project success, one small step at a time. Set reasonable timelines for goals and subgoals/steps to achieve the goal. Distribute work among team members depending on availability, e.g., define the scope of the paper, outline sections to be written, and distribute sections to different team members. Keep in mind that a common goal is a consensus about the path forward, not a competition for one side to convince the other (or take over the GAC), and not even necessarily a theoretical or scientific consensus. Ensure you have taken all viewpoints from the community into account by engaging the community early, e.g., in kick-off workshops, discussion forums, open reviews etc. Early feedback is usually rare in the research process; our experience showed that early feedback was extremely useful. Invite people to chime in. Some GACs might not be able to find a common goal, and that is ok. What remains important is a fruitful dialogue and to communicate the outcomes of this dialogue to the research community. Some GACs might make great progress in just defining and clarifying the controversy and agreeing on terminology for the first time. Others might end up identifying tutorial content or methodological approaches that could start to resolve a more established controversy. And others might come out with a solid foundation for a grant proposal, or even a plan for collecting new experimental data collected during the GAC, or new models built. Regardless of the specific outcome, the important aspect of GACs is that opposing camps work together to move the field forward. 

\textbf{Rule 3: Build your core GAC team well.} GAC teams can be fairly large and unwieldy. A crucial factor of success is to have a dedicated core team that drives the GAC. This allows other interested but maybe less available team members to be more in the ‘orbit’ and participate productively. Also, core team members will track progress, distribute tasks (on a volunteer basis), and hold each team member accountable with respect to their voluntary commitments. Thus, the core team should be highly motivated and energetic. Like for every collaboration, this works best when the core team has much mutual respect and compatible personalities. Early career researchers (ECRs) and junior scientists will have the strongest incentives to drive the GAC to a successful conclusion (see \nameref{sec:incentive}), and are also often poised to best drive the whole team towards cooperation since they may feel less committed to one theoretical ‘camp’. It is therefore recommended that ECRs play key roles in the core team. 

\textbf{Rule 4: Engage and incentivize broader community involvement.} This is a crucial ingredient of successful GACs, but it is also the hardest to achieve. The main hurdle is the potential lack of perceived incentives to engage in GACs. While the benefit of GACs for the field is clear, their benefits to individual researchers might need to be communicated more explicitly (e.g., the opportunity to become a driver of the field, develop collaborative grants, more publications, or increased visibility). There are also logistical hurdles for involvement that require technological solutions. For example, have a discussion forum (e.g., Slack, Discord), keep detailed meeting notes (e.g., Google Docs), and have a joint folder for materials (e.g., Google Drive) to facilitate access to information. Another opportunity for community engagement is to organize open online journal clubs. Community engagement will also grow with frequent communication on social media platforms, at conferences/workshops, in blog posts, and through journals (publications). Emphasize that community feedback and engagement are desired and specify the mechanisms through which such feedback can occur (e.g., OpenReview, commentary papers, workshop participation, etc). Finally, have a formal way for the GAC team to evaluate and utilize community feedback and engagement, but without diluting your goals or your core group. 

\textbf{Rule 5: Seek funding.} Use your GAC opinion piece and your ongoing collaboration to seek out funding mechanisms, and use this to find a way to continue the GAC into the future. This is as much about establishing a real collaboration as it is about keeping GAC members accountable and engaged. Project funding allows many researchers to justify their time commitments. It gives them credit in the current scientific environment. And it formalizes interactions with other GAC team members. There does not necessarily have to be large amounts of money, but even small funds can go a long way in achieving this engagement. Since ideally a GAC team is composed of some of the leaders of the field with the explicit goal to set aside differences and work out a jointly agreed upon way forward to make real progress in science, your GAC team funding proposal is likely highly competitive. Finally, your opinion paper could be the basis for such a funding proposal.

\subsection*{7 lessons for organizing and running a GAC program}

Organizing and running a successful GAC program can be challenging on many fronts. Below, we provide some lessons we learned from organizing the CCN GAC program in the hope that those lessons will be useful for others who wish to develop and launch their own version, whether conferences, institutions, or individual research groups.

\textbf{Lesson 1: Recognize and target changes in how GAC activities integrate with academic ‘currency’.} Why would anyone want to engage in a GAC? After all, it sounds like a lot of work. While that is true, there are also many benefits for the field and potentially for individual researchers. One way that organizers can incentivize the organization and continuation of GACs is by laying out avenues by which researchers can get credit in the current scientific environment. Specifically, can organizers…

\begin{enumerate}
    \item invite funding bodies to the GAC kick-off events in order to advertise promising research endeavors and potentially secure funding? 
    \item provide avenues for publication of opinion papers, e.g. through partnering with a journal for a special issue (like CCN has done with NBDT)?
    \item point out funding calls that might be able to support GAC teams? 
    \item provide a platform for continuing conversation with the community, e.g. through conference sessions? 
\end{enumerate}

Ultimately, anything that can be done to maximize the chances that team members get credit (in the traditional academic sense) for their involvement in GAC teams will maximize the chance for the successful continuation of the GAC teams past the initial position paper. 

\textbf{Lesson 2: More actively seek and maintain community involvement.} Continued community engagement is crucial to maintain energy and broaden the impact of the GACs. For example, our GAC proposal OpenReview was not very well populated. Organizers should thus work with GAC teams to develop multiple clear, front-facing, obvious opt-in ways that people can choose to participate or engage with a particular GAC, and to remove as many logistical barriers to doing so as possible. Different modalities could include:

\begin{itemize}
    \item Create mailing lists for each GAC of interested people, to allow the GAC program organizers as well as GAC leaders to reach out to interested people directly (email can clutter inboxes).
    \item Use Slack (or similar) communication platforms (people might go into information overload).
    \item Create a website and keep it up to date (requires checking the website and remembering to do it).
    \item Create a living document, such as a Wiki or Google Doc and keep it up to date (also requires the community to access and read these documents).
    \item Create an RSS feed/twitter timeline that could be pushed to people who want to sign up in their preferred modality.
    \item Solicit ``commentary'' papers to accompany GAC proposals and position papers, because these fall into the more traditional scientific credit system (as opposed to e.g., providing open reviews).
\end{itemize}

It is important to develop a way to regularly remind the community about the available resources and benefits of engaging with ongoing GACs. A scheduled reminder can help. 

\textbf{Lesson 3: Be clear about deadlines/timelines for GAC position papers (and other deliverables). }Establishing certain deliverables is important to provide guidance, structure, and a timeline to GAC teams. While those expectations should of course be reasonable (we are all busy people), it is also important to recognize that one size will not fit all GAC teams. Some teams might run into more difficulties progressing than others, some might need more time discussing, some might need more time engaging new community members, etc. Our experience has also shown that GAC teams often only realize certain issues when actively engaged in the writing process, which leads to the re-opening of discussions and thus delays. Timelines and expectations should thus be flexible. Nevertheless, timelines are important to set the pace and are great for momentum and morale. Deadlines energize projects. This is often a question of prioritization when being notoriously overcommitted. Regarding position papers for example, it is desirable to establish a short, but realistic timeline for those (CCN had a due date of ~6 months after the kick-off workshop). The main motivation is that such a position paper should be written when ideas are still fresh; the longer one waits, the more our memories will fail us. Thus, establishing due dates and sending out reminders is important, but it should be clear that these are suggested timelines and expectations rather than strict ones (assuming that there are no other external constraints, e.g., journal special issue due dates, etc).

\textbf{Lesson 4: Be clear about what the expectations are for the position paper. }Position papers are crucial components of the GAC process (see above). Therefore, it is also important for GAC program organizers to clearly lay out expectations and provide guidelines for these position papers. Beyond the timeline for the position papers (see above), the following (flexible) instructions help the GAC teams visualize what’s expected and needed.

Providing clarity on the structure of the position paper is important. Provide guidelines for the paper, and degree of flexibility/acceptable divergence from those guidelines. Typically this could for example be: general introduction, opposing viewpoints, proposed way forward, discussion. However, this ‘obvious’ structure of the position piece as a monolith may not accurately reflect everybody’s individual viewpoints. For example, a given sentence might be written and it might seem that everybody on the author list agrees with that sentence, even if they don’t, just based on the structure of the piece. One solution to this would be to adopt a signpost-type style clearly indicating whose position is laid out in a sentence, paragraph, or section, and to then also include alternative positions. A nice way to practically implement this could be through adopting a dialogue structure. After an introduction to establish the debate, written by all, a conversational flow of opposing viewpoints could present the arguments in something that feels more like a narrative. This has the clear advantage of allowing readers to follow the chain of argumentation from all points of view. 

Program organizers should also consider how to best facilitate the publication and ongoing public discussion of these opinion pieces. For example, we encourage GAC program organizers to negotiate a special issue in a journal that organizers (not GAC leaders/members) would edit. This mechanism allows full control by the program organizers over the content and process. Importantly, a bespoke special issue also allows for innovation in how the publication venue and process can engage public community interaction. For example, reviewers could be invited to publish a commentary on the opinion pieces they reviewed, similar to the rest of the community being invited to do the same. Maybe GAC teams should be encouraged to write commentaries and follow-up pieces on their own main position piece, to keep the development of the argument thread in a collated collection. As such, the position paper (within a special journal issue) becomes the start of a thread that tracks the field’s thinking over time. These mechanisms can also interface with existing credit assignment mechanisms (particularly peer-reviewed publications), providing concrete incentives to engage both GAC members and the wider community.

\textbf{Lesson 5: Different goals are okay.} As mentioned before, GAC teams can have different goals. For example, some GACs might make great progress in defining and clarifying the controversy and agreeing on terminology for the first time. Others might end up identifying tutorial content or methodological approaches that could start to resolve a more established controversy. Others might come out with a solid foundation for a grant proposal. Program organizers should encourage this diversity; explicit communication of different potential goals to GAC teams is important upfront to ensure that potential options and avenues of progress are clear to everyone. 

\textbf{Lesson 6: Develop a reward structure.} Motivation is important to all of us. But in science, playing into established reward structures and credit systems is at least equally important. Program organizers should thus explicitly lay out potential avenues for teams to get credit for the hard work they are doing in their GAC project. Potential options could include:

\begin{itemize}
    \item Papers. Organizers could provide an avenue for publication, e.g., negotiating a special GAC issue with a journal in the field.
    \item Code. Is there the potential for new code bases to be developed? If so, organizers could provide guidelines on how to make those the most useful for the field and/or how to disseminate them.
    \item Standards. Are GAC teams developing new standards for experimental design data collection or data analysis? Organizers could ensure there is an outlet for those methods to be distributed to the community, e.g., through mailing lists.
    \item Pilot funding. Organizers should consider seeking support from funding agencies for seed grants or sub-awards to be distributed to GAC teams (CCN GAC organizers are currently engaged in efforts towards this goal). In addition, program organizers should explicitly encourage GAC teams to write joint grant proposals, which are likely to be highly competitive.
    \item Conferences. Another good way to add value to the GAC teams’ work is to provide outlets at conferences for GAC teams to present their progress and advertise their initiatives. 
    \item Teaching. Are GAC teams developing new materials that require educating the community? Program organizers should think about innovative ways to encourage teams to develop high-quality teaching materials, e.g. by offering tutorial slots at conferences. 
\end{itemize}

\textbf{Lesson 7: Get granting agencies and journal editors involved.} An important way to ensure the continued impact and support of GAC teams is to invite representatives from granting agencies (e.g., program officers) and journal editors to not just attend GAC workshops, but to become actively involved in shaping the GACs. For example, granting agencies could present at GAC events to communicate what they are looking for regarding funding such projects. This would have multiple benefits, from educating funding agencies and editors about the GAC process, to providing an early opportunity for GAC teams to get formally recognized and supported by different stakeholders. Program officers could also sign up to be informed of new developments in a particular organization’s GAC process, suggest resources that might be particularly useful for those GAC members’ projects, or even coordinate to craft calls within cross-national collaboration frameworks that explicitly seek to develop partnerships between agencies. In fact, a GAC proposal at CCN is about 2 pages long, which is an ideal length for a white paper to be sent directly to a program officer for initial feedback on programmatic fit, according to many funding agencies’ practices.

\section*{Conclusion}

GACs are -- in our opinion -- tremendously beneficial endeavors for advancing science in a collaborative way. Here, we have outlined specific benefits and challenges encountered by GAC teams, and what ingredients we saw as being crucial for successful GACs. Some of the challenges encountered surely stem from the fact that GACs are a relatively new approach to science and not much has been documented yet regarding the best modus operandi. While establishing the best process is clearly a work in progress, we hope that our analysis and experience with the process will help other researchers see the tremendous benefits and a clear path forward on how to establish their own GACs.



\printbibliography

\end{document}